\documentclass[prl,aps,twocolumn,showpacs]{revtex4}
  \usepackage{epsfig}
  \usepackage{amssymb}
  \begin{document}
  \title{Causality in Propagation of a Pulse in a Nonlinear Dispersive Medium}
  \author{   G. S. Agarwal$^{1,2}$ and Tarak Nath Dey$^{1}$   }
  \affiliation{ $^1$ Physical Research Laboratory, Navrangpura,
  Ahmedabad, India \\
  $^2$ Department of Physics, Oklahoma State University, Stillwater,
  OK-74078}
  \date{\today}

  \begin{abstract}

  We investigate the causal propagation of the pulse
  through dispersive media by very precise numerical
  solution of the coupled Maxwell-Bloch equations
  without any approximations about the strength of the
  input field. We study full nonlinear behavior of the
  pulse propagation through solid state media like ruby
  and alexandrite. We have demonstrated that the
  information carried by the discontinuity, {\it i.e},
  front of the pulse, moves inside the media with
  velocity $c$ even though the peak of the pulse can
  travel either with sub-luminal or with super-luminal
  velocity. We extend the argument of Levi-Civita
  to prove that the discontinuity would
  travel with velocity $c$ even in a nonlinear medium.

  \end{abstract}

  \vspace*{-0.5cm} \maketitle

  The propagation of a pulse of electromagnetic radiation through a
  linear medium depends critically on the dispersive properties of
  the medium. Sommerfeld  and Brillouin investigated this problem in
  great detail \cite{Brillouin_book}. They discussed how the group
  velocity could be very different depending on whether one is
  working in the region of anomalous dispersion or normal
  dispersion. Some of these results were verified
  \cite{Chu_PRL_1982}. Sommerfeld and Brillouin also answered an
  important question - how information travels through the medium.
  Interest in this subject has been revived
  \cite{Hau_Nature_1999,Budker_PRL_1999,Kash_PRL_1999,tarak_PRA_2003}
  since by using external laser fields one can produce a desired
  dispersion \cite{Tewari_PRL_1986,Harris_PRL_1989}. Such techniques
  have, in fact, led to the realization of ultraslow light. Based on
  an earlier suggestion of Chiao \cite{Steinberg_PRA_1994}, Wang
  {\it et al.} demonstrated the production of super-luminal
  propagation \cite{Wang_Nature_2000}. The propagation of a pulse in
  a linear medium can also be understood from the interference of
  various Fourier components present in the input pulse
  \cite{Dogariu_Opt_2001}. Stenner {\it et al.} were able to
  demonstrate the causal behavior in the propagation of pulses by
  studying the propagation of discontinuities in  pulse shapes
  \cite{Stenner_Nature_2003}.

  An important question is - how the discontinuities travel in a
  nonlinear medium? Needless to say that pulses in a nonlinear
  medium have been studied quite extensively since the early work of
  Mc-Call and Hahn
  \cite{McCall_PRL_1967,Grobe_PRL_1994,Hioe_PRL_1994,Agarwal_PRA_2000}.
  In this letter, we examine propagation of discontinuities in a
  nonlinear medium. We choose  solid state materials like ruby and
  alexandrite. Bigelow {\it et al.} studied pulse propagation in
  such materials and showed the possibility of both sub-luminal and
  super-luminal propagation
  \cite{Bigelow_PRL_2003,Bigelow_SCI_2003}. These authors
  specifically concentrated on the nonlinear regime. We use our
  earlier theoretical models \cite{Agarwal_PRL_2004} to study the
  propagation of the front of a pulse and its discontinuities in
  materials like ruby and alexandrite. We show how the argument of
  Levi-Civita [Ref.1,p.38] can be extended  to prove that the front
  would travel with velocity $c$ even in a nonlinear medium. This is
  true despite the fact that the group velocity could be either
  sub-luminal or super-luminal.

\begin{figure}
  \begin{center}
  \includegraphics[scale=0.65]{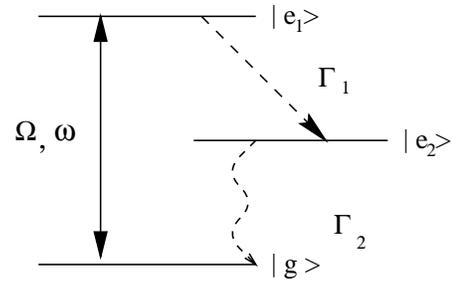}
  \caption{\label{Fig1} Schematic diagram of three level system for
  the ruby crystal.}
  \end{center}
  \end{figure}

\begin{figure}
  \centerline {
  \includegraphics[width=3 in]{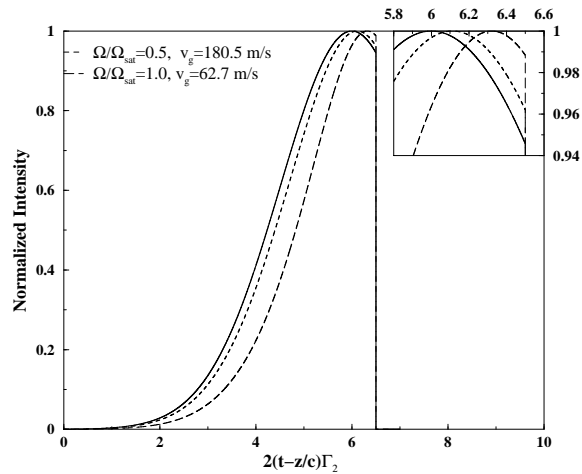}}
  \caption{\label{Fig2} The solid  curve depicts light pulse propagation
  through a free space with a velocity c. The long-dashed and dashed curves
  show the light pulse propagating through the medium at different intensities.
  The discontinuity of the pulse moves with a velocity c irrespective of medium or free space and
  obeys
  causality. The temporal
  width of the pulse is 20 ms and $1/2\Gamma_2=4.45$ms.}
  \end{figure}

  {\it Propagation of pulses with discontinuties in ruby}---In order to develop an
  understanding of causality in pulse propagation through ruby, we
  consider the model as shown in Fig~(\ref{Fig1}).
  The figure shows the energy level of chromium ion's (Cr$^{+3})$ in
  corundum (Al$_2$O$_3$) wherein we denote ground state level
  $^4$A$_2$ by $|g\rangle$, the $^{4}F_{2}$ absorption band by
  $|e_1\rangle$, and the doubly degenerate levels $2\bar{A},
  \bar{E}$ by $|e_2\rangle$. An intense field
  $\vec{E}(z,t)=\vec{\mathcal{E}}(z,t){\textrm exp}(ikz-i\omega t)
  +c.c.$ couples states $|g\rangle$ and $|e_1\rangle$ with Rabi
  frequency
  $\Omega(z,t)=2\vec{d}_{_{1g}}~\cdot~{\cal\vec{E}}(z,t)/\hbar$,
  where $\vec{d}_{_{1g}}$ is the dipole matrix element and
  $\vec{\mathcal E}$ is the slowly varying envelope of the intense
  electric field. Here we assume that the carrier frequency of
  intense field, $\omega$, is in resonance with the frequency of the
  $|g\rangle\longleftrightarrow|e_{_{1}}\rangle$ transition. The
  density matrix equations for the system, as shown in
  Fig.~(\ref{Fig1}), are given by \cite{Agarwal_PRL_2004}
  \begin{eqnarray}\label{density_ruby}
  \label{gg}\dot{\rho}_{_{gg}} &=& 2\Gamma_2\rho_{_{_{22}}} + i
  \frac{\Omega}{2}(\rho_{_{_{1g}}}-\rho_{_{_{g1}}})\nonumber\\
  \label{22}\dot{\rho}_{_{_{22}}} &=&
  2\Gamma_1\rho_{_{_{11}}}
  -2\Gamma_2\rho_{_{_{22}}}\nonumber\\
  \label{1g}\dot{\rho}_{_{_{1g}}} &=& -\Gamma_1\rho_{_{_{1g}}} + i
  \frac{\Omega}{2}(\rho_{_{gg}}-\rho_{_{_{11}}})\\
  \label{pop}\rho_{_{gg}}&+&\rho_{_{_{11}}}~+~~\rho_{_{_{22}}}=1,\nonumber
  \end{eqnarray}
  where $\rho_{_{_{ij}}}=\langle e_{_{i}}|\rho|e_{_{j}}\rangle$;
  $i,j=1,2$. Ruby has very strong relaxation effects where the
  dephasing rate $\Gamma_1$ is very large compared to population
  relaxation rate $\Gamma_2$ $(\Gamma_1\gg\Gamma_2)$. Typically
  $\Gamma_1\sim .15$ GHz and $\Gamma_2\sim 35$ Hz. Under the
  approximations $\Gamma_1\gg\Gamma_2,\Omega$, we obtain
  $\dot{\rho}_{1g}\sim0$ and then the evolution equation for the
  ground state population can be written in the form
  \begin{equation}\label{coher_approx}
  \dot{\rho}_{gg}=(1-\rho_{_{gg}}) -
  \frac{\Omega^2}{4\Gamma_1\Gamma_2}\rho_{_{_{gg}}}.
  \end{equation}
  In writing (\ref{coher_approx}) we have used the moving
  coordinates {\it i.e.,} the dot denotes the derivative with
  respect to $2(t-z/c)\Gamma_2$. Note that under
  $\Gamma_1\gg\Gamma_2,\Omega$ we can set $\rho_{_{11}}\approx 0$.
  In the slowly varying envelope approximation, we obtain the
  evolution of the Rabi frequency of an intense field as given by
  \begin{equation}\label{Max_field}
  \frac{\partial \tilde{\Omega}}{\partial z}= -
  \frac{\alpha_0}{2}\tilde{\Omega}\rho_{gg},~~~\tilde{\Omega}=\Omega/\Omega_{sat},
  \end{equation}
  where the coupling constant $\alpha_0 = 4{\cal
  N}\pi\omega|d_{_{1g}}|^2/c\hbar\Gamma_1$ and
  $\Omega_{sat}=\sqrt{4\Gamma_1\Gamma_2}$. In order to study the
  causality in the propagation of a pulse through ruby, we consider
  the shape of the pulse as incident on the input face of the medium
  given by
  \begin{eqnarray}\label{input_pulse}
  \tilde{\Omega}_{in}&=&\tilde{\Omega}^{0}e^{-[\tau/\sigma]^2}~~~\tau\leq T\nonumber\\
  &=&0~~~\tau > T.
  \end{eqnarray}
  A sharply vanishing amplitude gives rise to
  the discontinuity in the pulse. Here $\Omega^0$ is a real constant indicating
  the peak amplitude of the pulse and $\sigma$ is the temporal width of
  the input pulse. Once
  the shape of the input pulse is specified,
  the working equations (\ref{coher_approx}) and (\ref{Max_field})
  can be integrated numerically to investigate the evolution of the
  pulse. We consider all atoms to be initially in the state $|g\rangle$ at
  $\tau=0$.
  The length of the medium L and coupling constant
  $\alpha_0$ are chosen as 9 cm and 1.17 cm$^{-1}$ respectively.
  Fig.~(\ref{Fig2}) depicts the  propagation of the input pulse
  through ruby crystal which is known to exhibit normal dispersion.
  To prove that the discontinuity of the pulse moves inside ruby with
  a velocity $c$, numerical integration with
  high accuracy needs to be performed and the step size of the numerical integration
  to be taken of the order of $10^{-10}$ in the time domain.
  This is due to fact that the discontinuity of the pulse takes only 0.3 ns to
  traverse a path length of 9 cm in the medium. We found that the difference in time
  between the discontinuity of the pulse at output and input ends
  of the
  medium is $65\times10^{-9}$ in units of $1/2\Gamma_2$ by seeing the numerical
  output as given by table
  (\ref{tab:table1}),which is not well resolved in Fig.~(\ref{Fig2}).
  Therefore, the discontinuity of pulse moves inside the ruby crystal with velocity
  $c$, although, it is evident from Fig.~(\ref{Fig2}) that the peak of
  the pulse propagates with a subluminal velocity.
  This conforms with causality in the propagation of information
  through a nonlinear dispersive medium. We emphasize that in
  numerical simulations we have not used any kind of linearization
  on the nonlinear equations (\ref{coher_approx}) and
  (\ref{Max_field}). Note further that we work at intensities
  which are comparable to saturation intensities. The
  information travels in all cases with the velocity $c$ even
  though the peak position depends on the power of the pulse.

  \begin{table}
  \caption{\label{tab:table1}Numerical solution of the coupled
  Maxwell-Bloch equations for the input pulse with peak amplitude
  $\widetilde{\Omega}^{0}=1$ and discontinuity occurring at 6.5 in units
  of $1/2\Gamma_2$ are given in this table. The path
  length traversed by the pulse through vacuum as well as the medium
  is 9 cm. The difference in time for
  the discontinuity of the pulse to traverse the medium
  is $65\times10^{-9}$ in units of $1/2\Gamma_2$. All intensities are
  normalized to the peak values.}
  \begin{ruledtabular}
  \begin{tabular}{ccc}
  Retarded time&Output through&Output through\\
  $2(t-z/c)\Gamma_2$& vacuum &medium\\
  \hline
  6.4999999350&0.945844116&0.988440243\\
  6.4999999675&0.945843694&0.988440217\\
  6.5000000000&0.945843273&0.988440191\\
  6.5000000325&0.945842852&0.988440165\\
  6.5000000650&0.945842430&0.988440139\\
  6.5000000975&0.000000000&0.000000000\\
  \end{tabular}
  \end{ruledtabular}
  \end{table}

\begin{figure}
  \centerline {
  \includegraphics[width=3 in]{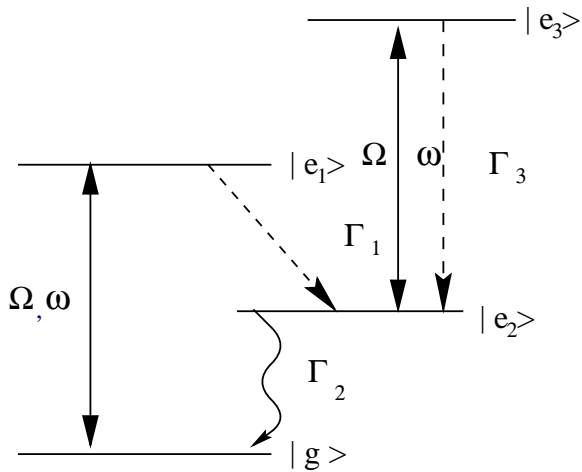}}
  \caption{Four level model for alexandrite crystal.}\label{Fig3}
  \end{figure}

\begin{figure}
  \centerline {
  \includegraphics[width=3.2 in]{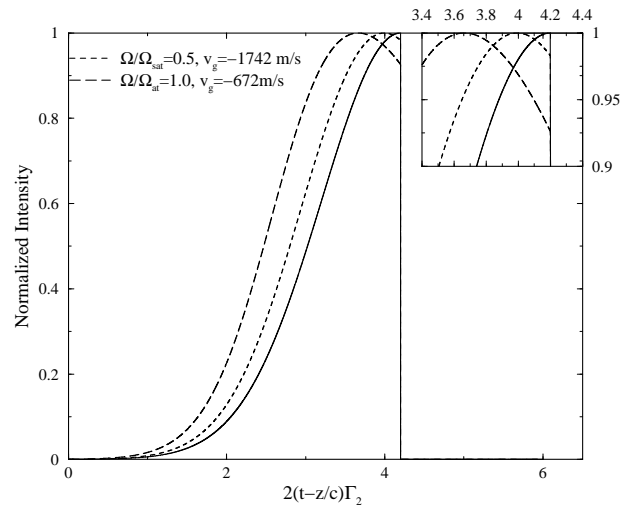}}
  \caption{\label{Fig4} The solid curve shows light pulse propagating
  at speed c. The dashed and long-dashed curves depict the propagation of the same pulse
  through an alexandrite crystal of length 9 cm with time delays $-51.7\mu$s and $-134\mu$s respectively,
  whereas the discontinuity of the pulse moves with a velocity c. The width $\sigma$
  and $1/2\Gamma_2$ are 500$\mu$s and $250\mu$s respectively.  }
  \end{figure}

  {\it Propagation of pulses with fronts in alexandrite}---
  To prove the causality of pulse propagation in a nonlinear anomalous dispersive medium, we consider
  alexandrite crystal where reverse saturation absorption is dominant. Reverse
  saturation absorption produces a narrow antihole in the
  susceptibility of the probe in the presence of a pump field which can
  lead to super-luminal propagation \cite{Bigelow_SCI_2003}. To
  investigate the propagation of the discontinuity of the pulse  through
  alexandrite, we refer to the ground state $^4$A$_2$
  by $|g\rangle$, the absorption bands $^4$T$_2$
  and $^4$T$_1$ by $|e_1\rangle$ and the level $^2$E by
  $|e_2\rangle$ as shown in Fig~(\ref{Fig3}).
  The intense pulse is defined by the electric field,
  $\vec{E}(z,t)=\vec{\mathcal E}(z,t)~e^{-i(\omega
  t-kz)}~+~c.c.$, which couples
  $|e_1\rangle\longleftrightarrow|g\rangle$, and also drives
  $|e_3\rangle\longleftrightarrow|e_2\rangle$. The density matrix
  equations for the model as shown in the Fig.~(\ref{Fig3}) are
  given by \cite{Agarwal_PRL_2004}
  \begin{eqnarray}\label{den_alex}
  \dot{\rho}_{_{_{gg}}} &=& 2\Gamma_2 \rho_{_{_{22}}} + i
  \Omega(\rho_{_{_{1g}}}-\rho_{_{_{g1}}})/2\nonumber\\
  \dot{\rho}_{_{_{22}}} &=&
  2\Gamma_1\rho_{_{_{11}}}
  -2\Gamma_2\rho_{_{_{22}}}+2\Gamma_3\rho_{_{_{33}}}+ i
  \Omega(\rho_{_{_{32}}}-\rho_{_{_{23}}})/2\nonumber\\
  \dot{\rho}_{_{_{33}}} &=& -2\Gamma_3 \rho_{_{_{33}}} + i
  \Omega(\rho_{_{_{23}}}-\rho_{_{_{32}}})/2\\
  \dot{\rho}_{_{_{32}}} &=& -\Gamma_3\rho_{_{_{32}}} + i
  \Omega(\rho_{_{_{22}}}-\rho_{_{_{33}}})/2\nonumber\\
  \dot{\rho}_{_{_{1g}}} &=& -\Gamma_1\rho_{_{_{1g}}} + i
  \Omega(\rho_{_{gg}}-\rho_{_{_{11}}})/2\nonumber\\
  \rho_{_{gg}}&+&\rho_{_{_{11}}}~~+~~\rho_{_{_{22}}}~~+~~\rho_{_{_{33}}}=1.\nonumber
  \end{eqnarray}
  Here we consider that the intense pulse is in
  resonance with $|e_1\rangle\longleftrightarrow|g\rangle$ and
  $|e_3\rangle\longleftrightarrow|e_2\rangle$, respectively.
  We can show that $\dot{\rho}_{_{32}}~{\textrm {and}}~\dot{\rho}_{_{1g}}$
  vanish when the approximations $\Gamma_1, \Gamma_3 \gg \Gamma_2,
  \Omega$; are taken into account. Under the same approximations,
  we obtain the equation for the evolution of ground
  state population as
  \begin{equation}
  \label{gg_alex} \dot{\rho}_{_{gg}}
  =(1-\rho_{_{_{gg}}})-|\tilde{\Omega}^2|\rho_{_{_{gg}}}.
  \end{equation}
  Note that $\rho_{_{_{11}}}$ \& $\rho_{_{_{33}}} \sim 0$ can be
  easily proved under $\Gamma_1, \Gamma_3 \gg \Gamma_2,
  \Omega$.
  Therefore, the evolution equation for the slowly varying Rabi frequency of the intense
  field is given by
  \begin{equation}\label{Max_alex}
  \frac{\partial \tilde{\Omega}}{\partial z}
  = -
  \frac{\alpha_0}{2}\tilde{\Omega}\rho_{_{_{gg}}}-\frac{\tilde{\alpha_0}}{2}
  \tilde{\Omega}(1-\rho_{_{_{gg}}})
  \end{equation}
  where $\alpha_0$ and $\tilde{\alpha}_0$ gives the saturation and reverse saturation, respectively.
  Shand {\it et al.} have shown that, for an excitation wavelength
  of 457 nm, the excited-state $(|e_2\rangle)$ absorption cross-section
  $(\sigma_2\sim4.05\times 10^{-20}$ cm$^2)$ exceeds that of the ground
  state $(|g\rangle)$ absorption cross-section $(\sigma_1\sim0.9\times 10^{-20}$ cm$^2)$ \cite{Shand_JAP_1981}.
  Following these experimental data, we estimate
  $(\tilde{\alpha}_0/\alpha_0)\sim 4 $. The numerical integration
  of coupled Maxwell-Bloch equations (\ref{gg_alex}) and (\ref{Max_alex}) gives the evolution
  of the input pulse as stated by eq.~(\ref{input_pulse}). Here we
  follow the same numerical integration procedure as stated in the
  case of ruby. We obtain the time for the discontinuity of the
  pulse to
  propagate through alexandrite as well as vacuum to be $12\times 10^{-7}$
  in units of $1/2\Gamma_2$ by checking the output of the numerical
  result (table (\ref{tab:table2})). This implies that the discontinuity of pulse moves
  inside the media with a velocity of light in free space $c$.
  We also found that the group velocity of the input pulse is in the range of
  superluminal velocity as illustrated in Fig.~(\ref{Fig4}). Therefore, we present a very
  precise numerical simulation that proves that the discontinuity in
  a pulse of a electromagnetic radiation cannot propagate faster
  than the velocity of light $c$ even though medium exhibits nonlinear anomalous
  dispersion. Consequently, the information carried by the
  discontinuity cannot  propagate with superluminal velocity.
  It should be borne in mind that the used intensities in the
  numerical simulation are comparable with the saturation
  intensities and the solution of the coupled equations
  (\ref{gg_alex}) and (\ref{Max_alex}) do not use any kind of
  perturbation. Hence, we have studied the full nonlinear causal behavior
  of pulses through alexandrite.

  \begin{table}
  \caption{\label{tab:table2}This table gives the result of the
  numerical integrations. The input pulse has peak amplitude
  $\widetilde{\Omega}^{0}=1$ with the discontinuity at 4.2 in units
  of $1/2\Gamma_2$. It shows that the time taken by
  the discontinuity of the pulse to propagate through medium as
  well as vacuum is $12\times10^{-7}$ in units of $1/2\Gamma_2$.}
  \begin{ruledtabular}
  \begin{tabular}{ccc}
  Retarded time&Output through&Output through\\
  $2(t-z/c)\Gamma_2$& vacuum &medium\\
  \hline
  4.2000000&0.9999999&0.92618654\\
  4.2000003&0.9999999&0.92618647\\
  4.2000006&0.9999999&0.92618639\\
  4.2000009&0.9999999&0.92618632\\
  4.2000012&1.0000000&0.92618624\\
  4.2000015&0.0000000&0.00000000\\
  \end{tabular}
  \end{ruledtabular}
  \end{table}

  Finally we show how the argument of Levi-Civita, as quoted in
  the book of Brioullion \cite{Brillouin_book}, can be extended
  to prove causality for propagation in a nonlinear medium. Let
  us consider a plane electromagnetic field say $\hat{x}$
  polarized and travelling in the $z$-direction so that the
  electric and magnetic vector have the form
  $\vec{E}=\hat{x}E(z,t); \vec{B}=\hat{y}B(z,t)$. The
  Maxwell equations then give us
  \begin{eqnarray}
  \label{Maxeq1}\frac{\partial E}{\partial z} &=&-
  \frac{1}{c}\frac{\partial
  B}{\partial t}\nonumber\\
  \label{Maxeq2}\frac{\partial B}{\partial z} &=&-
  \frac{1}{c}\frac{\partial E}{\partial t}-
  \frac{4\pi}{c}\frac{\partial {\mathcal P}}{\partial t}
  \end{eqnarray}
  where ${\mathcal P}$ is the nonlinear polarization of the medium. The
  medium is isotropic and hence the polarization is in $\hat{x}$
  direction and is a function of $z, t$. Let $e$ and $h$ be the
  discontinuities in $\partial E/\partial z$ and $\partial B/\partial
  z$ at boundary of the wavefront. Note that at the boundary of
  the wavefront $E$, $B$ and $\partial {\mathcal P}/\partial t$ are
  continuous. If $v$ represents the velocity with which the
  discontinuity moves in the direction of its normal then the
  identity compatibility relations imply that
  \begin{equation}\label{com1}
  {\textrm {discontinuity in}}~~\frac{\partial E}{\partial t} = -ve,
  \end{equation}
  and
  \begin{equation}\label{com2}
  {\textrm {discontinuity in}}~~\frac{\partial B}{\partial t} = -vh,
  \end{equation}
  This is seen from the equation
  $E(z+\Delta z, t+\Delta t)~-~E(z,t)=\Delta t~
  \partial E/\partial t~+~\Delta z~ \partial E/\partial z$.
  Using compatibility relations~(\ref{com1},\ref{com2}) and Maxwell
  eqs.~(\ref{Maxeq2}) we get $v=c$.

  In conclusion, we have established by both numerical simulations
  and analytical methods that the front travels with the velocity
  of light in free space even in a nonlinear medium.

  GSA thanks R. Boyd and M.S. Biegelow for preliminary data on
  their experiments.

  \end{document}